\documentclass[aps,preprint,preprintnumbers,amsmath,amssymb,nofootinbib]{revtex4}
\usepackage{amssymb}
\usepackage{amsmath}
\usepackage{bm}
\usepackage{graphicx}
\usepackage{epsfig}
\begin{document}
\title{Super exponential inflation from a dynamical foliation of a 5D vacuum state.}
\author{$^{1,2}$ Mauricio Bellini \footnote{ E-mail address:
mbellini@mdp.edu.ar, mbellini@conicet.gov.ar} }
\address{$^1$ Departamento de F\'isica, Facultad de Ciencias Exactas y
Naturales, Universidad Nacional de Mar del Plata, Funes 3350,
C.P. 7600, Mar del Plata, Argentina.\\  \\
$^2$ Instituto de Investigaciones F\'{\i}sicas de Mar del Plata (IFIMAR), \\
Consejo acional de Investigaciones Cient\'ificas y T\'ecnicas
(CONICET), Argentina.}

\begin{abstract}
We introduce super exponential inflation ($\omega < -1$) from a 5D
Riemann-flat canonical metric on which we make a dynamical
foliation. The resulting metric describes a super accelerated
expansion for the early universe well-known as super exponential
inflation that, for very large times, tends to an asymptotic de
Sitter (vacuum dominated) expansion. The scalar field fluctuations
are analyzed. The important result here obtained is that the
spectral index for energy density fluctuations is not scale
invariant, and for cosmological scales becomes $n_s(k<k_*) \simeq
1$. However, for astrophysical scales this spectrum changes to
negative values $n_s(k>k_*) <0 $.
\end{abstract}

\maketitle

\section{Introduction}

Observations of the anisotropy in the cosmic microwave background
radiation, are difficult to reconcile with a flat spectrum. For a
universe dominated by stable dark matter a flat spectrum is ruled
out altogether and even for a neutrino-dominated universe a
spectrum of the type $\delta \rho/\rho = (M/M_0)^{-s}$, with
$s>0$, is preferable to the flat spectrum (for which $s=0$).
Unfortunately, however, a quasi-exponential inflation with
$\dot{H} <0$ leads to a spectrum with $s <0$. This can be easily
understood intuitively larger mass scales correspond to
perturbations which crossed the horizon $1/H$ earlier during the
quasi-exponential phase of expansion, when $H$ was larger, which
implies a larger $\delta\rho/\rho$.

It is possible to construct models in which $\dot{H} >0$ by
starting with a theory in more than four dimensions\cite{pollock}.
In a higher-dimensional theory of gravity containing
higher-derivative terms and a cosmological constant, a period of
super-exponential inflation of the physical spacetime is possible,
during which the Hubble parameter H increases with time, as
discovered by Shafi and Wetterich\cite{SW}.

In this letter we explore this possibility from extra dimensions
using the Space-Time-Matter theory of gravity\cite{wo,we}, but
using a dynamical foliation $\psi\equiv \psi(t)$ of the noncompact
extra dimension $\psi$. This topic was explored previously by
Ponce de Le\'on\cite{pdl}. He showed that the FRW line element can
be "reinvented" on a dynamical four-dimensional hypersurface,
which is not orthogonal to the extra dimension, without any
internal contradiction. This hypersurface is selected by the
requirement of continuity of the metric and depends explicitly on
the evolution of the extra dimension. More recently, was
demonstrated that phantom\cite{pc} and warm
inflationary\cite{ruso} cosmological scenarios can be obtained
through this mechanism from a 5D Riemann-flat. In this letter
shall find that this mechanism can be the responsible for
existence of a super-exponential expansion of the early universe.

\section{Dynamical Foliation}

In this section we are interested to study the relativistic
dynamics of observers who moves on an 5D Ricci-flat spacetime,
such that the extra coordinate can be parameterized as a function
of an effective 4D spacetime.

We consider a 5D canonical metric
\begin{equation}\label{m}
dS^2 = g_{\mu\nu} (y^{\sigma},\psi) d y^{\mu} dy^{\nu} - d\psi^2.
\end{equation}
Here the 5D coordinates are orthogonal: $y \equiv
\{y^a\}$\footnote{Greek letters run from $0$ to $3$, and arabic
letters run from $0$ to $4$.}. The geodesic equations for an
relativistic observer are
\begin{equation}\label{geo}
\frac{dU^a}{dS} + \Gamma^a_{bc}\,\,U^b U^c =0,
\end{equation}
where $U^a = {dy^a\over dS}$ are the velocities and
$\Gamma^a_{bc}$ are the connections of (\ref{m}). Now we consider
a parametrization $\psi(x^{\alpha})$, where $x\equiv
\{x^{\alpha}\}$ are an orthogonal system of coordinates, such that
the effective line element (\ref{m}), now can be written as
\begin{equation}
dS^2 = h_{\alpha\beta}\,dx^{\alpha} dx^{\beta}.
\end{equation}
It is very important to notice that $S$ will be an invariant, so
that derivatives with respect to $S$ will be the same on 5D or 4D.
In other words, in this paper we shall consider that spacetime
lengths that remain unaltered when we move on an effective 4D
spacetime. The effective 4D tensor metric is $h_{\alpha\beta} =
{\partial y^a\over \partial x^{\alpha}} {\partial y^b \over
\partial x^{\beta}}\,g_{ab} = e^a_{\alpha} e^b_{\beta}\,g_{ab}$. Furthermore, since both coordinate systems,
$\{x^{\alpha}\}$ and $\{y^a\}$ are orthogonal and $S$ is an
invariant, hence we can define the inverse transformation $ g_{ab}
={\partial x^{\alpha}\over
\partial y^{a}} {\partial x^{\beta} \over
\partial y^{b}} \, h_{\alpha\beta}= \bar{e}^{\alpha}_a \bar{e}^{\beta}_{b}\, h_{\alpha\beta}$, such
that $e^a_{\alpha} \bar{e}^{\beta}_a = \delta^{\beta}_{\alpha}$.
In other words the condition of normalization for the velocities
of the observers is fulfilled in both the systems:
\begin{equation}
\left. g_{ab}\, U^a U^b \right|_{\psi(x^{\alpha})}=
h_{\alpha\beta} \, u^{\alpha} u^{\beta} =1,
\end{equation}
where we have used that $U^a \equiv \left(dy^a/dS\right) =
e^a_{\,\,\alpha} u^{\alpha} \equiv e^a_{\,\,\alpha}
\left(dx^{\alpha}/dS\right) $. If we replace this expression in
(\ref{geo}), we obtain
\begin{equation}
\frac{d\left(e^a_{\alpha} u^{\alpha}\right)}{dS} +
\Gamma^a_{\,\,bc} \left(e^b_{\,\,\alpha} e^c_{\,\,\beta}\right) \,
u^{\alpha} u^{\beta} =0,
\end{equation}
and hence
\begin{equation}
\bar{e}^{\phi}_a\,\frac{d}{dS}\left(e^a_{\,\,\alpha}
u^{\alpha}\right) + \Gamma^a_{\,\,bc} \left(e^b_{\,\,\alpha}
e^c_{\,\,\beta}\bar{e}^{\phi}_{\,\,a}\right) \, u^{\alpha}
u^{\beta} =0.
\end{equation}
Now we can use the following expression for the connection
transformation
\begin{equation}\label{co}
\bar{\Gamma}^{\phi}_{\alpha\beta} = \Gamma^a_{\,\,bc}
\left(e^b_{\,\,\alpha}
e^c_{\,\,\beta}\bar{e}^{\phi}_{\,\,a}\right)+\bar{e}^{\phi}_{\,\,a}
\,\,\frac{\partial}{\partial x^{\alpha}}
\left(e^a_{\,\,\beta}\right),
\end{equation}
and we obtain the geodesic equations for observers on the
effective 4D spacetime
\begin{equation}
\frac{d u^{\phi}}{dS} + \bar{\Gamma}^{\phi}_{\alpha\beta} \, u
^{\alpha} u^{\beta} = {\cal F}^{\phi},
\end{equation}
where the induced extra force on the effective 4D spacetime is
\begin{equation}\label{ef}
{\cal F}^{\phi}=\bar{e}^{\phi}_{\,\,a} \frac{\partial }{\partial
x^{\alpha}}\left(e^a_{\,\,\beta}\right) \,\,u^{\alpha} u^{\beta} -
u^{\alpha}\,\,\bar{e}^{\phi}_{\,\,a} \,\frac{d}{dS}\left(
e^a_{\,\,\alpha}\right).
\end{equation}
The extra force (\ref{ef}) can lead to the same result earlier
obtained in\cite{as}. In general, a nonzero ${\cal F}^{\phi}$
should be responsible for the violation of the equivalence
principle on the 4D hypersurface.

\subsection{Einstein equations for dynamical foliations from a 5D vacuum state}

Now we consider the Einstein equations on the 5D canonical
symmetrical metric
\begin{equation}
G_{ab} = -8\pi G\, T_{ab},
\end{equation}
where the Einstein tensor is given by $G_{ab} = R_{ab} -
\frac{1}{2} g_{ab} R$, $R_{ab}$ being the Ricci tensor, such that
the scalar of curvature is $R= g_{ab} R^{ab}$. Because we are
considering a 5D Ricci-flat metric, the Einstein tensor and the
Ricci scalar will be null. Using the transformation of the
previous section, we obtain that
\begin{equation}
\bar{G}_{\alpha\beta} = \bar{R}_{\alpha\beta} - \frac{1}{2}
h_{\alpha\beta} \bar{R} = -8\pi G \,\bar{T}_{\alpha\beta},
\end{equation}
where we have used respectively the transformations
\begin{eqnarray}
&& \bar{R}_{\alpha\beta} =  e^a_{\,\,\alpha} e^b_{\,\,\beta} \,R_{ab}, \\
&& \bar{R} = h_{\alpha\beta} \bar{R}^{\alpha\beta}, \\
&& \bar{T}_{\alpha\beta} = e^a_{\,\,\alpha} e^b_{\,\,\beta}
\,T_{ab},
\end{eqnarray}
for the effective 4D Ricci tensor, the scalar of curvature and the
energy momentum tensor.

\subsection{Energy Momentum tensor from a massless scalar field in
a 5D vacuum}

In order to make a complete description for the dynamics of the
scalar field, we shall consider its energy momentum tensor. In
order to describe a true 5D physical vacuum we shall consider that
the field is massless and there is absence of interaction on the
5D Ricci-flat manifold, so that
\begin{equation}
T^a_{\,\,b} = \Pi^a \Pi_b - g^a_{\,\,b} \,{\cal
L}\left[\varphi,\varphi_{,c}\right] ,
\end{equation}
where ${\cal L}\left[\varphi,\varphi_{,c}\right] = {1\over 2}
\varphi^a\varphi_a$ is the lagrangian density for a free and
massless scalar field on (\ref{m}) and the canonical momentum is
$\Pi^a = {\partial {\cal L}\over
\partial \varphi_{,a}}$. Notice that we are not considering interactions on the 5D vacuum,
because it is related to a physical vacuum in the sense that the
Einstein tensor is zero: $G^a_b =0$. The effective 4D energy
momentum tensor will be
\begin{equation}
\bar{T}_{\alpha\beta} = \left.e^a_{\,\,\alpha} \, e^b_{\,\,\beta}
\,T_{ab}\right|_{\psi(x^{\alpha})}.
\end{equation}
In other words, using the fact that ${\cal L}$ is an invariant it
is easy to demonstrate that
\begin{equation}
\bar{T}^{\alpha}_{\,\,\beta} = \bar{\Pi}^{\alpha}\,
\bar{\Pi}_{\beta} - h^{\alpha}_{\,\,\beta} {\cal L},
\end{equation}
where ${\cal L}$ is an invariant of the theory
\begin{equation}\label{ao}
{\cal L} = \frac{1}{2} \varphi^{,a} \varphi_{,a} = \frac{1}{2}
\left(e^a_{\alpha} \bar{\varphi}^{,\alpha}\right)
\left(\bar{e}^{\beta}_a \bar{\varphi}_{,\beta}\right) ,
\end{equation}
such that the Euler-Lagrange equations
\begin{equation}\label{bo}
\frac{\delta {\cal L}}{\delta \varphi} - \nabla_a \frac{\delta
{\cal L}}{\delta \varphi_{,a}}=0,
\end{equation}
describes the dynamics of the field $\varphi$ on the metric
(\ref{m}).

\subsection{D'Alambertian of a scalar field for dynamical foliations}

We are interested to study how is the effective 4D dynamics
obtained from a dynamic foliation of a 5D Ricci-flat canonical
metric. We consider a classical massless scalar field
$\varphi(y^a)$ on the metric (\ref{m}). The equation of motion
(\ref{bo}) for the Lagrangian density (\ref{ao}) is given by $\Box
\varphi = \nabla^a \varphi_{,a}=0$, where
\begin{equation}\label{dl}
\Box \varphi = g^{ab} \left(\varphi_{,a,b} - \Gamma^c_{ab} \,
\left[\varphi(y)\right]_{,c} \right).
\end{equation}
We can make the transformation from $y\equiv\{y^a\}$ to
$x\equiv\{x^{\alpha}\}$, with the parametrization
$\psi(x^{\alpha})$, where $x\equiv \{x^{\alpha}\}$. To simplify
the notation we shall use the notation
$\varphi(x)\equiv\bar{\varphi}$ and $\varphi(y)\equiv \varphi$,
such that
\begin{eqnarray}
&& \left.g^{ab} \varphi_{,a,b}\right|_{\psi(x^{\alpha})} =
 h^{\alpha\beta} \, \bar{\varphi}_{,\alpha,\beta}, \label{aa} \\
&& \left. g^{ab} \,\Gamma^c_{ab} \, \varphi_{,c}
\right|_{\psi(x^{\alpha})}= h^{\alpha\beta} \left[
\bar{\Gamma}^{\gamma}_{\alpha\beta} - \bar{e}^{\gamma}_{\,\,c}
\frac{\partial}{\partial x^{\alpha}}\left(e^c_{\,\,\beta}\right)
\right] \,\bar{\varphi}_{,\gamma}, \label{bb}
\end{eqnarray}
where in the last expression we have used the transformation law
for the connections (\ref{co}). Finally, using the expressions
(\ref{aa}) and (\ref{bb}) in (\ref{dl}), we obtain
\begin{equation}
\left.\Box \varphi\right|_{\psi(x^{\alpha})} =
\bar{\Box}\bar{\varphi} + h^{\alpha\beta}
\,\bar{e}^{\gamma}_{\,\,c} \frac{\partial}{\partial x^{\alpha}}
\left(e^c_{\,\,\beta}\right) \, \bar{\varphi}_{,\gamma},
\end{equation}
where $ \bar{\Box}\bar{\varphi}= h^{\alpha\beta}\,\left(
\bar{\varphi}_{,\alpha,\beta} -
\bar{\Gamma}^{\gamma}_{\alpha\beta} \,\bar{\varphi}_{,\gamma}
\right)$ is the effective 4D D'Alambertian on the orthogonal
coordinates system $x\equiv \{x^{\alpha}\}$ of the scalar field
$\bar{\varphi}(x^{\alpha})$ and the
$\bar{\Gamma}^{\gamma}_{\alpha\beta}$ are the Levi-Civita
connections on the effective 4D metric.

We shall require that
\begin{equation}\label{po}
\left.h^{\alpha\beta} \,\bar{e}^{\gamma}_{\,\,c}
\frac{\partial}{\partial x^{\alpha}} \left(e^{c}_{\,\,\beta}
\right) \, \varphi_{,\gamma}\right|_{\psi(x^{\alpha})} = \frac{d
}{d\,\bar{\varphi}} \,V(\bar{\varphi}).
\end{equation}
Because we are considering that $S$ and ${\cal L}$ remain
invariants on the effective 4D spacetime, only conservative
potentials will be used in (\ref{po}). In particular, the equation
of motion for a massive scalar field with imaginary mass ($i\,m$),
on the effective 4D spacetime is
\begin{equation}\label{ecc}
\bar{\Box} \bar{\varphi} - m^2\,\bar{\varphi} =0,
\end{equation}
which describes the motion of the scalar field
$\bar{\varphi}(x^{\alpha})$ on the 4D hypersurface with orthogonal
coordinates $x\equiv \left\{x^{\alpha}\right\}$. These
superluminal particles are known as tachyons\cite{as}.

\section{Super exponential inflation from a 5D canonical
metric}

In order to investigate an example of the earlier formalism, we
shall consider the Riemann-flat canonical metric\cite{LB}
\begin{equation}
dS^2 = \psi^2 dN^2 - \psi^2 e^{2N} dr^2 - d\psi^2,
\end{equation}
with the transformation: $\{y^a\} \rightarrow \{x^{\alpha}\}$
\begin{equation}
x^i = y^i \psi_0,  \qquad t = N \psi_0,
\end{equation}
and the dynamic foliation: $\psi \equiv \psi(t)$. The effective 4D
spacetime being described by the line element
\begin{equation}\label{met}
dS^2 = \left[\frac{\psi^2(t)}{\psi^2_0} - \dot\psi^2\right] dt^2
-\frac{\psi^2(t)}{\psi^2_0} \, e^{2\psi^{-1}_0 t} dR^2,
\end{equation}
where the dot denotes the derivative with respect to $t$ and
$\psi_0$ is some constant. In order to consider $t$ as a cosmic
time, one must require that
\begin{equation}
\frac{\psi^2(t)}{\psi^2_0} - \dot\psi^2 =1,
\end{equation}
so that the foliation is described by
\begin{equation}\label{ue}
\psi(t)= \psi_0\, \cosh{\left(t/\psi_0\right)}, \quad \rightarrow
\dot\psi(t) = \sinh{\left(t/\psi_0\right)}.
\end{equation}
Finally, the metric (\ref{met}), for a foliation (\ref{ue}) is
described by
\begin{equation}
dS^2 = dt^2 - \cosh{(t/\psi_0)}^2 \, e^{2\psi^{-1}_0 t}\, dR^2,
\end{equation}
which describes an 3D (flat) spatially isotropic universe in
expansion with a scale factor $a(t) = \cosh{\left(t/\psi_0\right)}
\, e^{\psi^{-1}_0 t}$, a scale factor $H(t)={\dot{a}\over a}$ and
a deceleration parameter $q=-{\ddot{a} a\over \dot{a}^2}$ given by
(for $H_0=2/\psi_0$)
\begin{eqnarray}
&& H(t) = \frac{H_0}{2} \left[ \tanh{\left(\frac{t H_0}{2}\right)}
+1\right], \\
&& q(t) = - \frac{2}{\tanh{\left(\frac{t H_0}{2}\right)} +1}.
\end{eqnarray}
Notice that $\dot{H} >0$. Furthermore the late time asymptotic
derivative the Hubble parameter and the deceleration parameter,
are
\begin{eqnarray}
&& \left.\dot{H}(t)\right|_{t\rightarrow \infty} \rightarrow H_0, \\
&& \left.{q}(t)\right|_{t\rightarrow \infty} \rightarrow -1,
\end{eqnarray}
which means that the universe describe an super exponential
inflationary period with and asymptotic de Sitter (vacuum
dominated) expansion.

On the other hand, the relevant components of the Einstein tensor,
are
\begin{eqnarray}
G^0_{\,\,0} &=& -\frac{3}{4} \frac{H^2_0 \left[ \sinh{(H_0 t/2)}
+ \cosh{(H_0 t/2)} \right]^2}{\cosh^2{(H_0 t/2)}}= -8\pi G \,\rho, \\
G^x_{\,\,x} &= & -\frac{H^2_0}{4} \frac{ \left[ \sinh{(H_0 t/2)} +
\cosh{(H_0 t/2)} \right] \left[ 5 \cosh{(H_0 t/2)} + \sinh{(H_0
t/2)} \right]}{\cosh^2{(H_0 t/2)}}=8\pi G \,p,
\end{eqnarray}
such that $G^x_{\,\,x}=G^y_{\,\,y}=G^z_{\,\,z}$, because the
isotropy of the space. The equation of state for this model of
super exponential inflation is
\begin{equation}
\frac{p}{\rho} = -\frac{1}{3} \frac{\left[ \sinh^2{(H_0 t/2)} + 5
\cosh^2{(H_0 t/2)} + 6 \sinh{(H_0 t/2)}\, \cosh{(H_0
t/2)}\right]}{\left[ \sinh^2{(H_0 t/2)} + \cosh^2{(H_0 t/2)} + 2\,
\sinh{(H_0 t/2)}\,\cosh{(H_0 t/2)}\right]}.
\end{equation}
Notice that its asymptotic value
\begin{equation}
\left.\frac{p}{\rho}\right|_{t\rightarrow \infty}\rightarrow -1,
\end{equation}
corresponds to a de Sitter exponential inflationary stage, with
cosmological constant
\begin{equation}
\Lambda_0 = 3 H^2_0 = \frac{12}{\psi^2_0}.
\end{equation}

\subsection{Evolution of $\bar{\varphi}$}

We consider the equation (\ref{po}). For a massive scalar field,
this equation can be rewritten as
\begin{equation}
\left.\ddot{\psi}\,
\frac{\partial\varphi}{\partial\psi}\right|_{\psi(t)} =
-m^2\,\bar{\varphi}(\vec{x},t),
\end{equation}
so that, using the fact that $\psi^2_0\,\ddot{\psi} =\psi$, we
obtain the solution $\varphi(\psi) =
\bar{\varphi}_0\,(\psi/\psi_0)^{-A}$, where $A=m^2 \psi^2_0 >0$ is
a dimensionless parameter. The equation of motion for the modes
$\bar{\varphi}_k(t)$ in eq. (\ref{ecc}), is
\begin{equation}\label{mod}
\ddot{\bar{\varphi}}_k + 3 H(t) \, \dot{\bar{\varphi}}_k + \left[
\frac{k^2}{a^2(t)} - m^2\right] {\bar{\varphi}}_k(t) = 0.
\end{equation}
The general solution is
\begin{eqnarray}
{\bar{\varphi}}_k(t) &=  & A_1
\left[x(t)+1\right]^{-\left(1+\sqrt{1-\left(\frac{2k}{H_0}\right)^2}\right)}
e^{\frac{H_0}{2} \left(-3 +
\sqrt{9+\left(\frac{2m}{H_0}\right)^2}\right)\, t} \, {\cal F}\left\{[a_1,b_1],[c_1],-x(t)\right\} \nonumber \\
& + &
B_1\,\left[x(t)+1\right]^{-\left(1+\sqrt{1-\left(\frac{2k}{H_0}\right)^2}\right)}
e^{-\frac{H_0}{2} \left(3 +
\sqrt{9+\left(\frac{2m}{H_0}\right)^2}\right)\, t} \, {\cal
F}\left\{[a_2,b_2],[c_2],-x(t)\right\}, \label{solu}
\end{eqnarray}
where ${\cal F}\left\{[a_i,b_i],[c_i],-x(t)\right\}$ is the
hypergeometric function with argument $x(t)=e^{-H_0 t}$, $i=1,2$,
and parameters
\begin{eqnarray}
a_{1} & = & -\sqrt{1-\left(\frac{2k}{H_0}\right)^2}+\frac{1}{2} -\frac{1}{2} \sqrt{9+\left(\frac{2m}{H_0}\right)^2} + \sqrt{\left(\frac{m}{H_0}\right)^2-4 \left(\frac{2k}{H_0}\right)^2} , \\
a_2  & = & -\sqrt{1-\left(\frac{2k}{H_0}\right)^2}+\frac{1}{2} +\frac{1}{2} \sqrt{9+\left(\frac{2m}{H_0}\right)^2} +  \sqrt{\left(\frac{m}{H_0}\right)^2-4 \left(\frac{2k}{H_0}\right)^2}, \\
b_1 & = & -\sqrt{1-\left(\frac{2k}{H_0}\right)^2}+\frac{1}{2} -\frac{1}{2} \sqrt{9+\left(\frac{2m}{H_0}\right)^2} -  \sqrt{\left(\frac{m}{H_0}\right)^2-4 \left(\frac{2k}{H_0}\right)^2}, \\
b_2 & = & -\sqrt{1-\left(\frac{2k}{H_0}\right)^2}+\frac{1}{2} +\frac{1}{2} \sqrt{9+\left(\frac{2m}{H_0}\right)^2} -  \sqrt{\left(\frac{m}{H_0}\right)^2-4 \left(\frac{2k}{H_0}\right)^2}, \\
c_1 & = & 1-\sqrt{9+\left(\frac{2m}{H_0}\right)^2} , \\
c_2 & = & 1+\sqrt{9+\left(\frac{2m}{H_0}\right)^2},
\end{eqnarray}
which we shall consider to be real.

\subsection{Fluctuations and spectrum}

\begin{figure*}
\includegraphics[height=16cm]{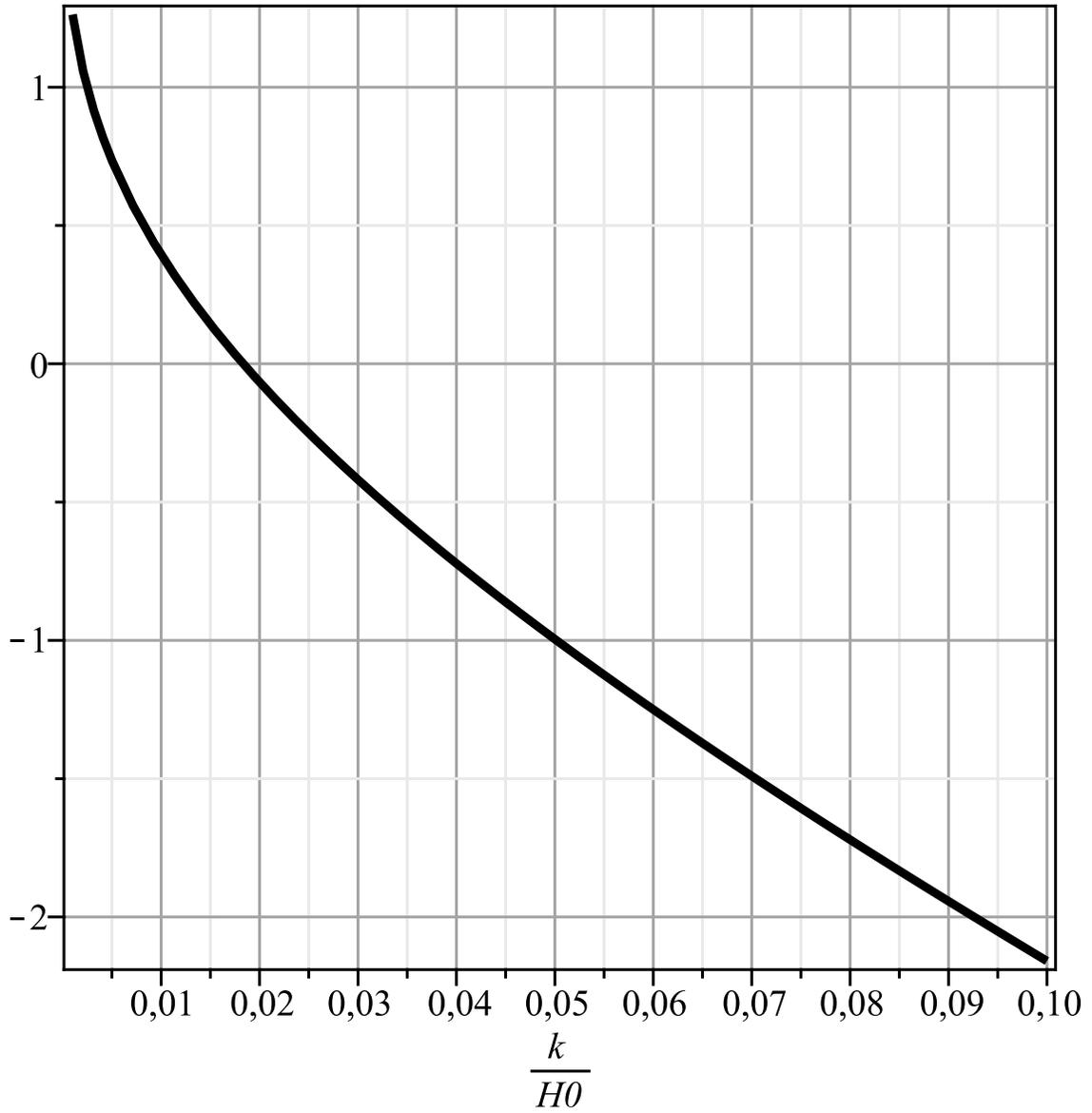}\caption{\label{fgi63} Spectral index $n_s(k)$ as a function of $k/H_0$. }
\end{figure*}

For large times the hypergeometric function is irrelevant: ${\cal
F}\left\{[a_i,b_i],[c_i],-x(t)\right\} \simeq 1$. Notice that the
exponential time dependent factor of the second term in
(\ref{solu}) decreases faster than the first one. Hence, for very
large time the second term of (\ref{solu}) can be neglected.
Furthermore, on cosmological scales the wavenumber $k$ is very
small: ($k/H_0 \ll 1$), so that one can make the approximation
$1+\sqrt{1- \left(\frac{2k}{H_0}\right)^2} \simeq 2- {1\over 2}
\left(\frac{2k}{H_0}\right)^2$. On the other hand, we shall
consider that the mass of the scalar field is very small: $m/ H_0
\ll 1$. Hence, if we chose $B_1=0$, we obtain that
\begin{equation}
\left.\bar{\varphi}_k(t) \bar{\varphi}^*_k(t)\right|_{t/\psi_0 \gg
1} \simeq A_1 A^*_1 \, e^{-H_0 \left[ 2-\left(\frac{2m^2}{9H^2_0}
+
 k^2\right)\right]t},
\end{equation}
and the squared $\varphi$-fluctuations are
\begin{equation}
\left< \varphi^2\right> =\frac{- A_1 A^*_1}{64 \pi^2 }
\frac{H_0}{t^{3/2}}\, e^{-{2t\over 9 H_0} \left(9 H^2 -
m^2\right)} \left\{ 4 t^{1/2} \epsilon k_0 e^{{4 \epsilon^2
k^2_0\over H_0} t} + \sqrt{\pi H_0} \, {\bf Erf}\left[{2\epsilon
k_0 \sqrt{t}\over \sqrt{H_0}}\right]\right\},
\end{equation}
where $\epsilon \simeq 10^{-3}$, ${\bf Erf}\left[x\right] =
{(2/\sqrt{\pi})} \int^x_0 e^{-x'^2} \,dx'$ is the error function
and $k_0(t)$ is the wavenumber that separates the horizon is
\begin{equation}
k_0(t)=a(t) \left[\frac{9}{4} H^2(t) + \frac{3}{2} \dot{H} +
m^2\right]^{1/2}.
\end{equation}
The spectrum for the squared $\varphi$-fluctuations corresponds
with a $k$-dependent spectral index $n_s(k)$
\begin{equation}
n_s(k) =3 - 2 \left({H_0 t_*\over {\rm ln}(k_*/H_0)}\right) \left[
1 - \left(\frac{m^2}{18 H^2_0} + \frac{k^2}{H^2_0}\right) \right],
\end{equation}
such that $t_*$ is defined as the time when super-inflation ends:
\begin{equation}
k_* = H_0\, e^{-H_0 t_*},
\end{equation}
and inflation begins. The spectral index $n_s(k)$, is
\begin{equation}
n_s(k) = 5 - 2 \left(\frac{m^2}{18 H^2_0} +
\frac{k^2}{H^2_0}\right).
\end{equation}
In the fig. (\ref{fgi63}) we have plotted $n_s$ for
$0.001<k/H_0<0.1$. We have used the values $H_0=6\times 10^{-9}\,
{\rm M_p}$ and $m=10^{-9}\,{\rm M_p}$, ${\rm M_p}=1.2\times
10^{19}\,{\rm GeV}$ being the Planckian mass. Notice that for
cosmological scales the spectral index is close to the unity, but
for astrophysical scales its value becomes negative. This is an
important result that agrees with observational date and it is not
possible to be recovered in an standard inflationary model for a
single scalar field.

\section{Final Comments}

Starting from a 5D canonical metric on which we make a dynamical
foliation, in this letter we have studied a model for super
exponential inflation. Along the super accelerated expansion
$\dot\omega >0$, so that for very large times the equation of
state tends to an asymptotic de Sitter (vacuum dominated)
expansion with $\left. \omega \right|_{t \rightarrow \infty}
\rightarrow -1$. We have obtained the effective 4D dynamics of the
scalar field, which drives the expansion of the universe, from a
dynamical foliation of a 5D Riemann-flat canonical metric, on
which this field is considered as a test massless field. However,
the scalar field acquires dynamics and (imaginary) mass on the
effective 4D FRW and can be considered as a tachyon field on this
4D hypersurface which was obtained from a dynamical foliation of
this space the 5D Riemann-flat spacetime.

The scalar field fluctuations on very large scales were analyzed.
The important result here obtained is that the spectral index for
energy density fluctuations ($\delta\rho / \rho \sim
<\varphi^2>$), are scale invariant for cosmological scales becomes
$n_s(k<k_*) \simeq 1$, but for more shorten scales this spectrum
changes to take negative values $n_s(k>k_*) <0$, until take values
close to $-2$, in agreement with observations.

\end{document}